\documentclass[10pt,pra,twocolumn,superscriptaddress,showpacs,floatfix]{revtex4}

\usepackage{graphicx}
\usepackage{amssymb}
\usepackage{times}
\usepackage{amsmath}

\begin{document}

\title{Coherent State Quantum Key Distribution Without Random Basis
Switching}

\author{Christian Weedbrook} \affiliation{Quantum Optics Group, Department
of Physics, Faculty of Science, Australian National University,
ACT 0200, Australia}\affiliation{Department of Physics, University
of Queensland, St Lucia, Queensland 4072, Australia}

\author{Andrew M. Lance} \affiliation{Quantum Optics Group, Department
of Physics, Faculty of Science, Australian National University,
ACT 0200, Australia}

\author{Warwick P. Bowen} \affiliation{Quantum Optics Group,
Department of Physics, Faculty of Science, Australian National
University, ACT 0200, Australia}

\author{Thomas Symul} \affiliation{Quantum Optics Group, Department of
Physics, Faculty of Science, Australian National University, ACT
0200, Australia}

\author{Timothy C. Ralph} \affiliation{Department of Physics, University of Queensland, St Lucia,
Queensland 4072, Australia}

\author{Ping Koy Lam} \affiliation{Quantum Optics Group, Department of
Physics, Faculty of Science, Australian National University, ACT
0200, Australia}

\date{\today}

\begin{abstract}

The random switching of measurement bases is commonly assumed to
be a necessary step of quantum key distribution protocols. In this
paper we show that switching is not required for coherent state
continuous variable quantum key distribution. We show this via the
\textit{no-switching protocol} which results in higher information
rates and a simpler experimental setup. We propose an optimal
eavesdropping attack against this protocol, for individual
Gaussian attacks, and we investigate and compare the no-switching
protocol applied to the original BB84 scheme.

\end{abstract}

\pacs{03.67.Dd, 42.50.Dv, 89.70.+c}

\maketitle

\section{Introduction}
Quantum key distribution (QKD) \cite{Wiesner,BB84,QKD} allows two
people, Alice and Bob, to communicate in secret, where the laws of
quantum physics insure the total privacy of their communication. A
secret key is generated by Alice transmitting quantum states to
Bob, who performs measurements on the received states. An
eavesdropper, Eve, who actively attacks the quantum channel, will
disturb the quantum system and hence be detected. A passive
eavesdropper, however, whose attack simulates the quantum channel
will remain undetected, but the maximum information obtainable by
this attack is known and can be bounded. Additionally Eve cannot
intercept and perfectly copy the quantum states as a consequence
of the no-cloning theorem of quantum information \cite{WZ82}.
After they have distilled a secret key, Alice and Bob can use this
key to communicate secret information over a classical
communication channel.

The first QKD protocol, known as the BB84 protocol, uses single
randomly polarized photon states \cite{BB84}. Here Alice prepares
and sends a random ensemble of single photon states over a quantum
channel to Bob. Bob then measures the states by randomly switching
between two non-commuting measurement bases - a compulsory step to
insure the security of the protocol. Any loss to the environment
or noise on the channel is attributed to Eve, who is only limited
in her attack by the laws of physics. Later Alice and Bob produce
a sifted key by discarding results where their bases are not the
same. Alice and Bob then release a part of their raw key to test
for channel transmission and errors in their correlated bit
string. Reconciliation protocols \cite{Mau93} are employed in
order to correct any errors between Alice and Bob's correlated
key. Finally privacy amplification \cite{Ben95} is used to reduce
Eve's knowledge of the key to a negligible amount. Once a secret
key has been generated it is used as a
 one-time pad \cite{Ver26} to encrypt the message. The
absolute security of the BB84 protocol has been proven in
\cite{May98,Lo99,Sho00}. Other single photon schemes proposed
include the EPR protocol \cite{Eke91} and the B92 protocol
\cite{Ben92}.

QKD using continuous variables was introduced in 1999 \cite{ral00}
as an alternative to the original single photon schemes.
Continuous variable \cite{Bra04} QKD offer the advantages of
higher detection efficiencies, compatibility with current
technologies and faster communication speeds. In 2000, continuous
variable QKD using squeezed states \cite{hil00} and EPR
correlations \cite{Rei00} were proposed. Further work included
using squeezed \cite{Cer01} and coherent states \cite{G&G3dB} that
generated Gaussian keys against individual Gaussian attacks. All
of these protocols were originally only thought to be secure for
line losses less than $50\%$ or 3dB of noise. However, it has been
shown that this limit can be overcome by using either reverse
reconciliation \cite{G&GReverse} or post-selection \cite{Sil02}
techniques. Reverse reconciliation involves Alice (and Eve)
estimating what states Bob has measured rather than the usual way
of Bob (and Eve) trying to determine what states Alice has sent.
In this sense the flow of classical information is in the reverse
direction. The second protocol, post-selection, can also tolerate
higher losses by Alice and Bob carefully selecting information for
which they have an advantage over Eve. The unconditional security
of continuous variable QKD has been proven for squeezed state
protocols \cite{Got01} and Gaussian modulated coherent states
using homodyne detection \cite{Ibl04}. Collective attacks using
reverse reconciliation and their unconditional security was also
discussed in \cite{Gro05}. Continuous variable QKD has also been
experimentally demonstrated in \cite{Gro03,Lor04,Lan05}.

The random switching of measurement bases by Bob has been a
fundamental step in QKD protocols using both single photon states
and continuous variables. However recently we introduced a new
coherent state continuous variable QKD protocol, against
individual Gaussian attacks, that does not require switching. This
new protocol, known as the {\it no-switching protocol}, involves
Bob measuring both bases simultaneously. This was shown to offer
higher information rates along with a simpler experimental setup
than protocols that use switching \cite{WLBS+04}. Since then the
no-switching protocol has lead to various research, both
theoretical \cite{Fil05,Gro05} and experimental \cite{Lan05}. It
was also shown in \cite{Gro05} to be secure against collective
eavesdropping attacks. In this paper we will expand on our
analysis of the no-switching protocol and consider a more thorough
eavesdropping attack. We will also discuss an equivalent
no-switching protocol for discrete variables, in particular the
BB84 protocol.

This paper is organized as follows. In Section II we discuss the
no-switching protocol for coherent state continuous variable QKD.
Section III discusses and compares two physical realizations of
possible eavesdropping attacks. In Section IV we discuss an
equivalent no-switching protocol for discrete variables. Section V
concludes.


\section{The No-Switching Protocol}

\subsection{Notation}

Before leading into the steps of the no-switching protocol we will
first briefly discuss the nomenclature used throughout this paper.
Quantum states that we consider in this paper can be described
using the state vector notation $|a\rangle$. These quantum states
can be described using the boson field annihilation operator
$\hat{a}$ which can be expressed in terms
of the quadrature field operators as %
\begin{eqnarray}\label{a}
\hat{a} = \frac{1}{2} (\hat{X}^{+} + i \hat{X}^{-})
\end{eqnarray}
where $\hat{X}^{+}$ and $\hat{X}^{-}$ are the amplitude (+) and
phase (-) quadrature operators respectively. The annihilation
operator is not measurable in itself as it is non-Hermitian.
However, we can express the real and imaginary parts of
Eq.(\ref{a}) as
\begin{eqnarray}
\hat{X}^{+} &=& \hat{a}^{\dag} + \hat{a}\\
\hat{X}^{-} &=& i(\hat{a}^{\dag} - \hat{a})
\end{eqnarray}
which are Hermitian and therefore measurable. These operators can
be expressed in terms of a steady state and a fluctuating
component as $\hat{X}^{\pm} = \langle \hat{X}^{\pm} \rangle +
\delta \hat{X}^{\pm}$ with quadrature variances defined as
$V^{\pm} = \langle (\delta \hat{X}^{\pm})^2 \rangle$. In this
paper all operators representing the amplitude and phase
observables are denoted by a hat symbol.

\subsection{No-Switching Protocol}

In the original coherent state QKD protocols \cite{Cer01,G&G3dB}
Alice first prepares a displaced vacuum state that will be sent to
Bob. This is achieved by choosing two real random numbers
$\mathcal{S}^{+}$ and $\mathcal{S}^{-}$ from a Gaussian
probability distribution with zero mean $\langle
\mathcal{S}^{\pm}\rangle = 0$ and a variance of
$V_{\mathcal{S}}^{\pm} = \langle (\mathcal{S}^{\pm})^2 \rangle$.
She then displaces the amplitude and phase quadratures of the
coherent state by $\mathcal{S}^{+}$ and $\mathcal{S}^{-}$
respectively. The displaced coherent state can be represented by
the state vector $|\mathcal{S}^{+} + i \mathcal{S}^{-} \rangle$.
This state has corresponding operators $\hat{X}_{A}^{\pm}$
associated with the amplitude and phase observables of the quantum
state. So typically we can express these operators and
corresponding quadrature variances $V_{A}^{\pm}$ as
\begin{eqnarray}\label{def Xa}
\hat{X}_{A}^{\pm} &=& \mathcal{S}^{\pm} + \hat{N}_{A}^{\pm}\\
V_{A}^{\pm} &=& V_{\mathcal S}^{\pm} + 1
\end{eqnarray}
where $\hat{N}_{A}^{\pm}$ is the operator associated with the
amplitude and phase of the initial vacuum state $|0\rangle$, which
has a normalized variance $ \langle (\hat{N}_{A}^{\pm})^2 \rangle
= 1$. Alice transmits this coherent state to Bob through a quantum
channel with a channel transmission efficiency $\eta$. The losses
in the channel couples in channel noise, which have corresponding
quadrature operators denoted as $\hat{X}_{N}^{\pm}$ (and with
corresponding quadrature variances $V_{N}^{\pm}$). Therefore the
states that arrive at Bob's station can be described by the
quadrature operators and corresponding variances
\begin{eqnarray}
\hat{X}_{B'}^{\pm} &=& \sqrt{\eta}\hat{X}_{A}^{\pm} +
\sqrt{1-\eta}\hat{X}_{N}\\
V_{B'}^{\pm} &=& \eta V_{A}^{\pm} + (1 - \eta)V_{N}^{\pm}
\end{eqnarray}
where the superscript ' indicates states entering Bob's station.
Once Bob receives these states he does not randomly switch between
measurement quadratures, instead he simultaneously measures both
the amplitude and phase quadratures of the state via a 50/50
beamsplitter and a pair of homodyne detectors. We denote Bob's
quadrature measurements by $\hat{X}_{B}^{\pm}$ with a quadrature
variance of $V_{B}^{\pm}$, which can be expressed as
\begin{eqnarray}\label{Bob_state}
\hat{X}_{B}^{\pm} &=& \frac{1}{\sqrt2}(\sqrt{\eta}
\hat{X}_{A}^{\pm} + \sqrt{1 - \eta} \hat{X}_{N}^{\pm} +
\hat{N}_{B}^{\pm})\\\label{Bob_Variance} V_{B}^{\pm} &=&
\frac{1}{2}(\eta V_{A}^{\pm} + (1 - \eta)V_{N}^{\pm} + 1).
\end{eqnarray}
\begin{figure}[!ht]
\begin{center}
\includegraphics[width=8cm]{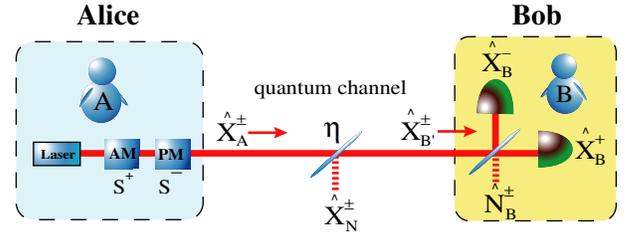}
\caption{Schematic of a coherent state continuous variable QKD
scheme using the no-switching protocol. AM: amplitude modulator,
PM: phase modulator; $\mathcal{S}^{\pm}$: random Gaussian numbers;
$\hat{X}_{A}^{\pm}$: Alice's prepared state; $\eta$: channel
transmission; $\hat{X}_{N}^{\pm}$: channel noise;
$\hat{X}_{B'}^{\pm}$: describes the states entering Bob's station;
$\hat{N}_{B}^{\pm}$: Bob's vacuum noise; $\hat{X}_{B}$: describes
the states Bob measures after the
beamsplitter.}\label{Noswitching_Schematic}
\end{center}
\end{figure}
where $\hat{N}_{B}^{\pm}$ is the vacuum noise entering into Bob's
$50/50$ beamsplitter with a variance $\langle
(\hat{N}_{B}^{\pm})^2\rangle =1$. Figure
\ref{Noswitching_Schematic} gives a schematic of the no-switching
coherent state QKD protocol. As a point of interest it has been
shown \cite{CRD04} that simultaneously measuring both quadratures
is equivalent to an optimal quantum cloner for continuous variable
Gaussian states, i.e. a cloning machine that produces two quantum
clones with an optimal fidelity of $\mathcal{F}=2/3$. So the
no-switching protocol can also be thought of as an optimal cloning
protocol where Bob (and, as we shall see, Eve) optimally clones
their respective quantum states.

A consequence of the no-switching protocol is that vacuum noise is
introduced into Bob's measurements via the 50/50 beamsplitter at
his station. In the following sections we analyze the effect of
this extra quanta of noise introduced into Bob's quadrature
measurements in terms of secret key rate and the overall security
of the protocol. We will show that it is still secure against
individual attacks and it results in higher information rates as a
result of obtaining two simultaneous streams of information from
both quadratures, instead of the usual one quadrature measurement.

\subsection{Mutual Information}

In analyzing QKD protocols we are inevitably concerned with the
net mutual information \cite{Shannon,Nie00} between Alice and Bob
in the presence of Eve, i.e. the rate at which a secret key can be
generated by Alice and Bob. In this paper we consider that Alice
and Bob use the reverse reconciliation protocol to generate a
secret key \cite{G&GReverse}. For the reverse reconciliation
protocol the net information rate can be written as
\begin{eqnarray}\label{secret_key_rate}
\Delta I = I(B:A) - I(B:E)
\end{eqnarray}
where $I(B:A)$ is the mutual information between Bob and Alice and
similarly between Bob and Eve, $I(B:E)$. We can define these
quantities as
\begin{eqnarray}\label{mutual_information1}
I(B:A) &=& H(B) - H(B|A)\\\label{mutual_information2} I(B:E) &=&
H(B) - H(B|E)
\end{eqnarray}
where $H(B)$ is Bob's Shannon entropy and $H(B|A)$ and $H(B|E)$
are Alice's and Eve's conditional entropies relative to Bob's
measurement \cite{Nie00} respectively. The conditional entropy is
a measure of how uncertain Alice and Eve are, on average, about
Bob's measurement result. These quantities can physically be
thought of as noise, due to the quantum channel and the intrinsic
quantum noise of Alice's state. For the no-switching protocol,
Alice encodes independent information onto both quadratures of a
coherent state, which Bob then simultaneously measures.
Subsequently we can describe the mutual information rate between
Alice and Bob as the sum of the quadrature information rates as
\begin{eqnarray}
\Delta I = \Delta I^{+} + \Delta I^{-}
\end{eqnarray}
where $\Delta I^{+}$ is the information rate for the amplitude
quadrature and $\Delta I^{-}$ the information rate for the phase
quadrature. By substituting
Eqs.(\ref{mutual_information1},\ref{mutual_information2}) into
Eq.(\ref{secret_key_rate}) and assuming symmetry for both
quadratures, we end up with the information rate for the
no-switching protocol given by
\begin{eqnarray}\label{info_SQM1}
\Delta I = 2 [ H(B|E) - H(B|A) ]
\end{eqnarray}
Hence, in order to determine the final information rate, we need
to determine both Alice's and Eve's conditional entropies.

\subsection{Conditional Variances}

We can express Alice's and Eve's conditional entropies of Bob's
quadrature measurements in terms of conditional variances
\cite{Shannon}
\begin{eqnarray}\label{Alice_Conditional_Entropy}
H(B|A) = \frac{1}{2} {\rm
log}_{2}(V_{A|B})\\\label{Eve_Conditional_Entropy} H(B|E)
=\frac{1}{2} {\rm log}_{2}(V_{E|B})
\end{eqnarray}
where $V_{A|B}$ and $V_{E|B}$ are Alice's and Eve's conditional
variances relative to Bob's measurement respectively \cite{GLP98}.
The conditional variance can be thought of as the uncertainty in
Alice's and Eve's estimates of Bob's quadrature measurement
result. In general the conditional variance of X given the event Y
can be written as
\begin{eqnarray}\label{DefGeneral_ConditionalVariance}
V_{X|Y} = var(X|Y) = \min_{g}\langle (Y - g X)^2\rangle
\end{eqnarray}
where $g$ is an optimal gain that minimizes the conditional
variance. Therefore, the total information rate for the
no-switching protocol given in Eq.(\ref{info_SQM1}), assuming
symmetry of both quadratures, can be written in a simpler form in
terms of conditional variances as
\begin{eqnarray}\label{info_SQM2}
\Delta I = {\rm log}_{2}\Big(\frac{V_{E|B}}{V_{A|B}}\Big)
\end{eqnarray}

\subsection{Alice's Conditional Variance}

For the no-switching protocol Alice's conditional variance of
Bob's measurements is defined as
\begin{eqnarray}\label{DefConditionalVariance}
V_{A|B}^{\pm} = \min_{g_{A}^{\pm}} \langle (\hat{X}_{B}^{\pm} -
g_{A}^{\pm} {\mathcal S}^{\pm})^2\rangle
\end{eqnarray}
where $\hat{X}_{B}^{\pm}$ is Bob's quadrature measurement given by
Eq.~(\ref{Bob_state}), $\mathcal{S}^{\pm}$ is the quadrature
displacement of Alice's prepared state and $g_{\rm A}^{\pm}$ is an
experimental gain or Alice's best estimate at what Bob has
measured. This gain is then optimized to give a minimum
conditional variance. The minimum gain is given by
$g_{A}^{\pm}=\langle{\mathcal{S}^{\pm}}\hat{X}_{B}^{\pm}\rangle/
\langle{\mathcal{S}^{\pm}}^{2}\rangle$ which is then substituted
into Eq.~(\ref{DefConditionalVariance}) to give a conditional
variance of
\begin{eqnarray}\label{Alice_Conditional_Variance}
V_{A|B}^{\pm} &=& V_{B}^{\pm}-\frac{\langle
\mathcal{S}^{\pm}X_{B}^{\pm}\rangle^{2}}{V_{\mathcal{S}}^{\pm}}
\end{eqnarray}
We now calculate Alice's conditional variances for the
no-switching protocol. To calculate the conditional variances we
consider a more general protocol, where Alice can transmit to Bob
displaced squeezed states instead of coherent states. This
scenario leads to the best possible correlation between Alice and
Bob for a particular quadrature measurement. In this case, the
quadrature variance of the states prepared by Alice are given by
\begin{eqnarray}\label{VA_with_sqz}
V_{A}^{\pm}=V_{{\mathcal{S}}}^{\pm}+V_{{sqz}}^{\pm}
\end{eqnarray}
where $V_{sqz}^{\pm}$ is the quadrature variance of the squeezed
states prepared by Alice. To ensure that the quadrature variances
of Alice's transmitted state remains symmetric with variances
$V_{A}^{\pm}$, we require that the maximum amount of squeezing is
limited by the variance of Alice's transmitted states. This is
given by the following inequality
\begin{eqnarray}\label{Vsqz_geq_Va}
V_{\rm sqz}^{\mp} \geq \frac{1}{V_{A}^{\pm}}
\end{eqnarray}
Substituting Eq.(\ref{Bob_Variance}) into
Eq.(\ref{Alice_Conditional_Variance}), with $\langle
{\mathcal{S}^{\pm}X_{B}^{\pm}}\rangle = \sqrt{\eta /2}
V_{\mathcal{S}}^{\pm}$, we can calculate Alice's conditional
variance to be
\begin{eqnarray}\label{AlicesCondVar}
V_{A|B}^{\pm} &=&\frac{1}{2}\Big(\eta V_{\rm
sqz}^{\pm}+(1-\eta)V_{N}^{\pm}+1\Big)
\end{eqnarray}
where we have used the fact that $V_{\mathcal{S}}^{\pm} =
V_{A}^{\pm} - V_{sqz}^{\pm}$ from Eq.(\ref{VA_with_sqz}). We point
out that in the no-switching protocol Alice does not actually use
squeezed states but rather coherent states. So eventually we will
set $V_{sqz}^{\pm} = 1$. We only consider that Alice uses
squeezing for our analysis in order to give lower bounds for the
quadrature conditional variances.

\subsection{Uncertainty Relations}

Before we explicitly calculate Eve's conditional variance we first
derive a general relationship between Alice's and Eve's
conditional variances. We will then use this relation to bound
Eve's minimum conditional variance and hence Eve's maximum mutual
information with Bob. To calculate a relationship between Alice's
and Eve's conditional variances of Bob's measurement,
$V_{E|B}^{\mp}$ and $V_{A|B}^{\pm}$, we define the operators that
denote Alice's and Eve's inference of Bob's measurement {\it
before his beamsplitter}, expressed as
\begin{eqnarray}
\hat{X}_{E|B'}^{\pm}=\hat{X}_{B'}^{\pm}- g_{E}^{\pm}\hat{X}_{E}^{\pm}\\
\hat{X}_{A|B'}^{\mp}=\hat{X}_{B'}^{\mp}- g_{A}^{\mp}
\mathcal{S}^{\mp}
\end{eqnarray}
where $g_{E}^{\pm} \hat{X}_{E}^{\pm}$ and $g_{A}^{\mp}
\mathcal{S}^{\mp}$ are Alice's and Eve's optimal estimates with
optimal gains, $g_{E}^{\pm}$ and $g_{A}^{\mp}$. Finding the
commutator of the above two equations, and using the fact that
different Hilbert spaces commute, we find that
\begin{eqnarray}
[\hat{X}_{E|B'}^{+},\hat{X}_{A|B'}^{-}]=[\hat{X}_{B'}^{+},\hat{X}_{B'}^{-}]=2i
\end{eqnarray}
This leads to \cite{Sak94} the joint Heisenberg
uncertainty relation
\begin{equation}\label{VarUncertainty}
V_{E|B'}^{\pm} V_{A|B'}^{\mp}\geq 1
\end{equation}
Therefore, there is a limit to what Alice and Eve can know
simultaneously about what Bob has measured. Once again an
important notational point is the superscript $'$. Whenever this
is used it implies that the equations are only dealing with the
states before Bob's beamsplitter. From the above inequality it is
possible to determine the maximum information Eve can obtain about
the state in terms of Alice's conditional variances
$V_{A|B}^{\pm}$.

\subsection{Eve's Conditional Variance}

We now calculate Eve's minimum conditional variance for an attack
which is only limited by the joint Heisenberg uncertainty relation
given in Eq.~(\ref{VarUncertainty}) and Alice's conditional
variance in Eq.~(\ref{AlicesCondVar}). To find a lower bound on
Eve's conditional variance, we first consider her inference of
Bob's state prior to the 50/50 beamsplitter in Bob's station. As
with Alice's conditional variance, Eve's conditional variance is
given by
\begin{eqnarray}
V_{E|B'}^{\pm}=\min_{g_{\rm E}^{\pm}} \langle(\hat{X}_{B'}^{\pm}
-g_{E}^{'\pm}\hat{X}_{E}^{\pm})^2 \rangle
\end{eqnarray}
where $\hat{X}_{B'}^{\pm}$ (defined in Eq.~(\ref{XB_switching}))
is the quadrature of Bob's state that could be measured prior to
the beamsplitter and the associated gain $g_{E}^{'\pm}$. Eve's
measurement variance after the beamsplitter conditioned on Bob's
measurement ($V_{E|B}^{\pm}$) can be expressed in terms of the
conditional variance before the beamsplitter ($V_{E|B'}^{\pm}$) as
\begin{eqnarray}\label{beforeAndAfterBeamSplitter}
\nonumber V_{E|B}^{\pm}&=&\left
\langle\left(\hat{X}_{B}^{\pm}-g_{E}^{\pm}\hat{X}_{E}^{\pm}\right)^2
\right \rangle\\\nonumber &=&\left \langle
\left(\frac{1}{\sqrt2}(\hat{X}_{B}^{\pm}+\hat{N}_{B}^{\pm})-g_{E}^{\pm}\hat{X}_{E}^{\pm}\right)^{2}
\right \rangle\\\nonumber &=&\frac{1}{2}\left \langle
\left(\hat{X}_{B}^{\pm}-\sqrt2
g_{E}^{\pm}\hat{X}_{E}^{\pm}\right)^{2} \right
\rangle+\frac{1}{2}\\\nonumber &=&\frac{1}{2}\left \langle
\left(\hat{X}_{B}^{\pm}-g_{E}^{'\pm}\hat{X}_{E}^{\pm}\right)^{2}
\right
\rangle+\frac{1}{2}\\
&=&\frac{1}{2}\Big(V_{E|B'}^{\pm}+1\Big)
\end{eqnarray}
%
\begin{figure}[!ht]
\begin{center}
\includegraphics[width=8cm]{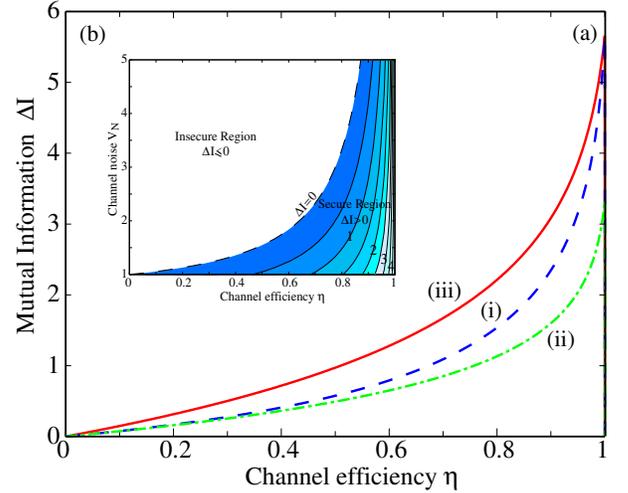}
\caption{(a) Net information rates for the no-switching (i)
and switching (ii) protocols as a function of channel efficiency;
where $V_{N}^{\pm}=1$ and $V_{A}^{\pm}=100$. Two eavesdropping
attacks are given by the same solid curve (iii). (b) Contour plot
of the information rate for the no-switching protocol as a
function of channel efficiency and channel noise.}
\label{FigureInformationRates}
\end{center}
\end{figure}
where we have used the fact that Eve has no access to the
beamsplitter in Bob's station, and therefore has no knowledge of
the vacuum entering through it. The uncertainty relation
Eq.(\ref{VarUncertainty}) tells us that there is a limit to what
Alice and Eve can simultaneously know about what Bob has measured,
i.e. $V_{E|B^{'}}^{\mp} \geq 1/V_{A|B^{'}}^{\pm}$. We can now
determine Alice's conditional variance of Bob, if Bob were to
directly measure a single quadrature of his state before the 50/50
beamsplitter $V_{\rm A|B'}^{\pm}$. The derivation of $V_{\rm
A|B'}^{\pm}$ is based on the derivation given in \cite{G&GReverse}
which goes as follows. We can use
Eq.(\ref{Alice_Conditional_Variance}) with the following equations
\begin{eqnarray}
{\mathcal {S}}^{\pm} &=& \hat{X}_{A}^{\pm} - \hat{N}_{A}^{\pm}\\
V_{\mathcal{S}}^{\pm} &=& V_{A}^{\pm} -
V_{sqz}^{\pm}\\\label{XB_switching}
\hat{X}_{B'}^{\pm} &=& \sqrt{\eta} \hat{X}_{A}^{\pm} + \sqrt{1 - \eta} \hat{X}_{N}^{\pm}\\
V_{B'}^{\pm} &=& \eta V_{A}^{\pm} + (1 - \eta)V_{N}^{\pm}\\
\langle \mathcal{S}^{\pm} \hat{X}_{B'}^{\pm}\rangle &=& \sqrt\eta
V_{\mathcal{S}}^{\pm}
\end{eqnarray}
The above equations are the same as the no-switching equations,
e.g. Eqs.(\ref{Bob_state},\ref{Bob_Variance}), except for the
$1/\sqrt2$ and $\hat{N}_{B}^{\pm}$ that are due to Bob's
simultaneous quadrature measurements. Alice's conditional variance
using reverse reconciliation {\it{with}} switching is given by
\begin{eqnarray}
V_{A|B'}^{\pm} \geq V_{A|B' min}^{\pm} = (\eta/
V_{A}^{\pm}+(1-\eta)V_{N}^{\pm})
\end{eqnarray}
Using this minimum value with Eq.(\ref{VarUncertainty}) we can
calculate Eve's conditional variance when switching is used
\begin{eqnarray}\label{VEB' Bound}
V_{E|B^{'}}^{\pm} \geq (\eta/V_{A}^{\pm} +
(1-\eta)V_{N}^{\pm})^{-1}.
\end{eqnarray}
Again we emphasize that we have assumed that Eve can
simultaneously measure both the amplitude and phase quadratures of
her ancilla (or measuring) state without paying a ``quantum
duty''. This is in fact an unphysical assumption that allows Eve
more information than what she is entitled to. We calculate a
lower bound on Eve's conditional variance for the no-switching
protocol by substituting Eq.~(\ref{VEB' Bound}) into
Eq.~(\ref{beforeAndAfterBeamSplitter})
\begin{equation}\label{EvesMinCondVar}
V_{\rm E|B}^{\pm} \geq
\frac{1}{2}\Big(\Big(\frac{\eta}{V_{A}^{\pm}}+(1-\eta)V_{N}^{\pm}\Big)^{-1}+1\Big).
\end{equation}

\subsection{Secret Key Rate}

We can now determine the final secret key rate for the
no-switching protocol by substituting Alice's and Eve's
conditional variances from Eqs.~(\ref{AlicesCondVar},
\ref{EvesMinCondVar}) (and symmetrize both quadratures) into
Eq.(\ref{info_SQM2}). The final secret key rate is then given by
\begin{eqnarray}\label{SecretKeyRateHeterodyne}
\Delta {\rm I} &\geq&\log_{2}
\Big(\frac{(\frac{\eta}{V_{A}}+(1-\eta)V_{N})^{-1}
+1}{\eta+(1-\eta)V_{N}+1} \Big)
\end{eqnarray}
where we have set $V_{sqz}^{\pm} = 1$ to indicate that we are
using coherent states. Figure~\ref{FigureInformationRates}(b)
shows a plot of Eq.~(\ref{SecretKeyRateHeterodyne}) for varying
channel transmissions and varying channel noise. We see that it is
completely secure for vacuum noise and as the noise is increased
the insecure region gets larger (as is the case for all QKD
protocols). Figure~\ref{FigureInformationRates}(a) plots the
information rate of Eq.~(\ref{SecretKeyRateHeterodyne}) (for
$V_{N}^{\pm} = 1$ and $V_{A}^{\pm} = 100$) against the channel
transmission (dashed line). We can now compare the no-switching
protocol (dashed line) to the switching protocol (dot dashed line)
of \cite{G&GReverse}. Figure~\ref{FigureInformationRates} shows
that the no-switching protocol has a higher information rate than
the switching protocol for all channel transmission losses.

We now turn our attention to physical implementations of
eavesdropping attacks against the no-switching protocol and
compare these attacks to the bound derived in
Eq.~(\ref{SecretKeyRateHeterodyne}).

\section{Eavesdropping Attack}

\begin{figure}[!ht]
\begin{center}
\includegraphics[width=8cm]{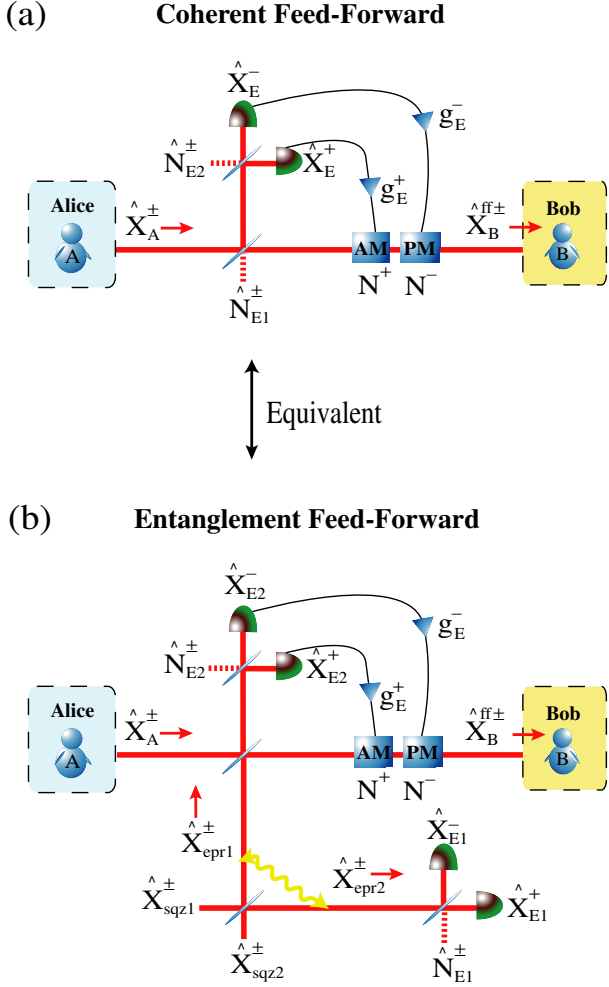}
\caption{Schematic of two possible eavesdropping attacks: (a)
Coherent feed-forward attack and (b) entanglement feed-forward
attack.}\label{Figure_Eve_Attack}
\end{center}
\end{figure}

In the previous section we derived an upper bound on Eve's maximum
information for the the no-switching protocol. In this section we
put a lower bound on Eve's information by investigating a physical
eavesdropping attack. A feed-forward scheme was discussed in
\cite{WLBS+04} as a possible eavesdropping attack to the
no-switching protocol for varying channel noise (i.e. not just
vacuum noise). Here Eve used a beamsplitter to gather information
from the quantum channel (see Fig.~\ref{Figure_Eve_Attack}a). She
then measured both quadratures simultaneously and then feed
forward these altered states onto Bob. We denote this attack as
the {\it coherent feed-forward attack}. We will now consider a
more sophisticated attack that incorporates additional
entanglement. We show that by giving Eve additional resources,
e.g. the use of entanglement, that she gets no additional
information compared to the coherent feed-forward attack. We then
discuss reasons for thinking that the coherent feed-forward attack
might be optimal for individual Gaussian attacks.

\subsection{Entanglement Feed-Forward Attack}

Figure \ref{Figure_Eve_Attack}b shows a schematic of the
entanglement feed-forward attack. This attack goes as follows: Eve
creates two Einstein-Podolsky-Rosen (EPR) \cite{Ein35} entangled
beams by interfering two squeezed beams $\hat{X}_{sqz1}^{\pm}$ and
$\hat{X}_{sqz2}^{\pm}$ on a 50/50 beamsplitter. The quadratures of
the entangled beams are described by the following operators
\begin{eqnarray}
\hat{X}_{epr1}^{\pm} &=& \frac{1}{\sqrt2} (\hat{X}_{sqz1}^{\pm} + \hat{X}_{sqz2}^{\pm})\\
\hat{X}_{epr2}^{\pm} &=& \frac{1}{\sqrt2} (\hat{X}_{sqz1}^{\pm} -
\hat{X}_{sqz2}^{\pm})
\end{eqnarray}
Eve retains the second of the entangled beams
$\hat{X}_{epr2}^{\pm}$, which she simultaneously measures both
quadratures via a 50/50 beamsplitter and two perfect homodyne
detectors. We assume that Eve has quantum memory and that she
performs these measurements only after Bob has receives the
states. These resulting quadrature operators and corresponding
variances are given by
\begin{eqnarray}
\hat{X}_{E1}^{\pm} &=& \frac{(\hat{X}_{sqz1}^{\pm} -
\hat{X}_{sqz2}^{\pm})
/\sqrt2 - \hat{N}_{E1}^{\pm}}{\sqrt2}\\
V_{E1}^{\pm} &=& \frac{(V_{sqz1}^{\pm} + V_{sqz2}^{\pm} + 2)}{4}
\end{eqnarray}
where $\hat{N}_{E1}^{\pm}$ is the vacuum noise entering into Eve's
first beamsplitter with $\langle (\hat{N}_{E1}^{\pm})^2 \rangle
=1$. Eve then injects the entangled beam $\hat{X}_{epr1}^{\pm}$
into a beamsplitter with a transmission efficiency $\epsilon$ to
simulate channel losses. Eve then simultaneously measures both
quadratures of the output beam of the beamsplitter, with
quadrature operators and corresponding variances expressed as
\begin{eqnarray}\nonumber
\hat{X}_{E2}^{\pm} &=& \Big(\sqrt{\frac{1-\epsilon}{2}}
\hspace{1mm} \hat{X}_{A}^{\pm}+\frac{\sqrt\epsilon}{2}\hspace{1mm}
(\hat{X}_{sqz1}^{\pm} +
\hat{X}_{sqz2}^{\pm}) + \frac{\hat{N}_{E2}^{\pm}}{\sqrt2} \Big)\\
V_{\rm E2}^{\pm}&=& ((1-\epsilon)V_{\rm
A}^{\pm}+\frac{\epsilon}{2}(V_{sqz1}^{\pm} + V_{sqz2}^{\pm})+1)/2
\end{eqnarray}
where $\hat{N}_{E2}^{\pm}$ is the vacuum noise entering into Eve's
second beamsplitter with $\langle (\hat{N}_{E2}^{\pm})^2 \rangle =
1$. Using lossless electro-optic feed-forward techniques, Eve
transfers the measured photocurrents back onto the quantum channel
with some gain $g_{\rm E}^{\pm}$. The gain of this feed-forward
must be chosen to ensure that the magnitude of the signal detected
by Bob remains invariant. So for an arbitrary channel transmission
$\eta$, Bob would be expecting a signal described by the
quadrature operator $\sqrt\eta \hspace{1mm} \hat{X}_{A}^{\pm}$ due
to losses in the quantum channel. Now Eve knows that she is
sending Bob a signal of $\sqrt\epsilon \hspace{1mm}
\hat{X}_{A}^{\pm}+g_{E}^{\pm} \sqrt{(1-\epsilon)/2}\hspace{1mm}
\hat{X}_{A}^{\pm}$. Therefore she wants $\sqrt\eta \hspace{1mm}
\hat{X}_{A}^{\pm}$ to equal $\sqrt\epsilon \hspace{1mm}
\hat{X}_{A}^{\pm}+g_{E}^{\pm} \sqrt{(1-\epsilon)/2}\hspace{1mm}
\hat{X}_{A}^{\pm}$. This leads to a gain of
\begin{eqnarray}
g_{E}^{\pm} =
\frac{\sqrt2(\sqrt\eta-\sqrt\epsilon)}{\sqrt{1-\epsilon}}
\end{eqnarray}
We point out that this is the same gain as given for the coherent
feed-forward attack \cite{WLBS+04}. Now because Eve has reduced
the transmission of the quantum channel, she will need to add
additional Gaussian noise $\mathcal{N}^{\pm}$ (with variance
$V_{\mathcal{N}}^{\pm}$) to remain undetected (which she can
subsequently infer out of her estimate of Bob's quadrature
measurement with arbitrary precision). Once Bob receives these
altered quantum states from Eve, he then measures both quadratures
simultaneously
\begin{eqnarray}\label{BobVarforEveAttack}\nonumber
\hat{X}_{B}^{\rm ff ^{\pm}} &=&[\sqrt{\epsilon} \hat{X}_{A}^{\pm} - \sqrt{(1-\epsilon)/2}(\hat{X}_{sqz1}^{\pm} + \hat{X}_{sqz2}^{\pm})\\
&+& \hat{N}_{B}^{\pm} + \mathcal{N}^{\pm}+ g(\sqrt{1 - \epsilon}
\hat{X}_{A}^{\pm}\\\nonumber &+& \sqrt{\epsilon
/2}(\hat{X}_{sqz1}^{\pm} + \hat{X}_{sqz2}^{\pm})+
\hat{N}_{E2}^{\pm})\sqrt2]\sqrt2
\end{eqnarray}
This has a corresponding variance of
\begin{eqnarray}\nonumber
 V_{B}^{\rm ff ^{\pm}} &=&
\eta V_{A}^{\pm}/2 +
V_{\mathcal{N}}^{\pm}/2 - [2 + V_{sqz1}^{\pm} + V_{sqz2}^{\pm}\\
&-& 2\sqrt{\eta \epsilon}(2 + V_{sqz1}^{\pm} + V_{sqz2}^{\pm}) + 2
\eta\\\nonumber &+& \eta \epsilon (V_{sqz1}^{\pm} +
V_{sqz2}^{\pm})]/(4 \epsilon -4)
\end{eqnarray}
We are now in a position to calculate Eve's conditional variance
$V_{E1,E2|B}^{\rm ff \pm}$ for this entanglement feed-forward
attack, as a function of the beamsplitter transmission $\epsilon$.
Once we have calculated this conditional variance we can then
proceed as before to determine the secret key rate. The
conditional variance in this case will be a tripartite conditional
variance defined as
\begin{eqnarray}
V_{E1,E2|B}^{\rm ff \pm} = \langle (\hat{X}_{B}^{\pm} -
g_{1}^{\pm}\hat{X}_{E1}^{\pm} - g_{2}^{\pm} \hat{X}_{E2}^{\pm})^2
\rangle
\end{eqnarray}
as we need to accommodate both of Eve's measurements in her
estimate of Bob's quadrature measurements. Minimizing the two
gains, we have
\begin{eqnarray}\label{DefEveCondVar}\nonumber
V_{E1,E2|B}^{\pm} &=& V_{B}^{\pm} - \frac{V_{E1}^{\pm}\langle
\hat{X}_{B}^{\pm}\hat{X}_{E2}^{\pm} \rangle^2 +
V_{E2}^{\pm} \langle \hat{X}_{B}^{\pm}\hat{X}_{E1}^{\pm} \rangle^2}{V_{E1}^{\pm}V_{E2}^{\pm} - \langle \hat{X}_{E1}^{\pm}\hat{X}_{E2}^{\pm} \rangle ^2}\\
&+& \frac{2\langle \hat{X}_{B}^{\pm}\hat{X}_{E1}^{\pm}\rangle
\langle \hat{X}_{B}^{\pm}\hat{X}_{E2}^{\pm}\rangle \langle
\hat{X}_{E1}^{\pm}\hat{X}_{E2}^{\pm}
\rangle}{V_{E1}^{\pm}V_{E2}^{\pm} - \langle
\hat{X}_{E1}^{\pm}\hat{X}_{E2}^{\pm} \rangle ^2}
\end{eqnarray}
The above equation is a function of the two channel transmissions
$\eta$ and $\epsilon$. To calculate one in terms of the other we
need to consider how much information Eve is allowed to have
before she is detected by Alice and Bob. This is related by the
following inequality
\begin{eqnarray}\nonumber
(1-\eta)V^{\pm}_{N}&\geq& \Big((V_{sqz1}^{\pm} + V_{sqz2}^{\pm} + 2\epsilon)\\
&-& 2\sqrt{\epsilon \eta}(2 + V_{sqz1}^{\pm} +
V_{sqz2}^{\pm}\\\nonumber &+& \eta (2 + \epsilon (V_{sqz1}^{\pm} +
V_{sqz2}^{\pm})))\Big)/2(1 - \epsilon)
\end{eqnarray}
where the left hand side is the amount of noise for an arbitrary
quantum channel. This noise must be greater than or equal to the
total noise Eve has put onto the channel (as can be seen from
Eq.~(\ref{BobVarforEveAttack})). We numerically minimize
$V_{E1,E2|B}^{{\rm ff \pm}}$ for all $\epsilon$ between
$\epsilon_{\rm min}$ and $\epsilon_{\rm max}$. Alice's conditional
variance remains the same as before (i.e.
Eq.(\ref{AlicesCondVar})) as Eve has applied the correct gain. Now
that we have Eve's and Alice's conditional variances we can use
Eq.(\ref{info_SQM2}) to numerically calculate the secret key rate.
We find that the secret key rate for the entanglement feed-forward
attack is the same as the secret key rate for the coherent
feed-forward attack.

In Fig.~\ref{FigureInformationRates} we plot both the entanglement
feed-forward attack and the coherent feed-forward attack (both the
same top solid line). We conjecture that this higher bound is in
fact the optimal bound for the no-switching protocol against
individual attacks. In principle the best Eve can do is to
optimally clone the states she intercepted from Alice and then
send these cloned copies back onto Bob - which is exactly what the
coherent feed-forward attack does. Also by giving Eve more
resources, such as entanglement, does not in any way give her an
information advantage. This suggests we have found an optimal
information bound.


\section{BB84 and the No-Switching Protocol}

As we have seen, the no-switching protocol works successfully in
the continuous variable regime using coherent states. It
eliminates the need to randomly switch bases resulting in
simplicity and higher information rates. We now ask the question:
can the no-switching protocol be applied to the discrete variable
regime and give the same benefits? To determine if this happens we
consider an equivalent no-switching protocol for single photon
states. This involves introducing a quantum cloning machine in
Bob's station that allows him to imperfectly clone (due to the
no-cloning theorem \cite{WZ82}) what Alice has sent into two sets
of copies. Bob can then measure both polarization bases
simultaneously. This cloning machine will negate the need for Bob
to randomly switch measurement bases. As a note, there have been
other discrete variable proposals that do not rely on the random
switching of measurement bases but rather the random switching of
state manipulation, such as the ``ping pong" protocol
\cite{Bos02}.

\subsection{How the No-Switching Protocol for BB84 Works}

\begin{figure}[!ht]
\begin{center}
\includegraphics[width=8cm]{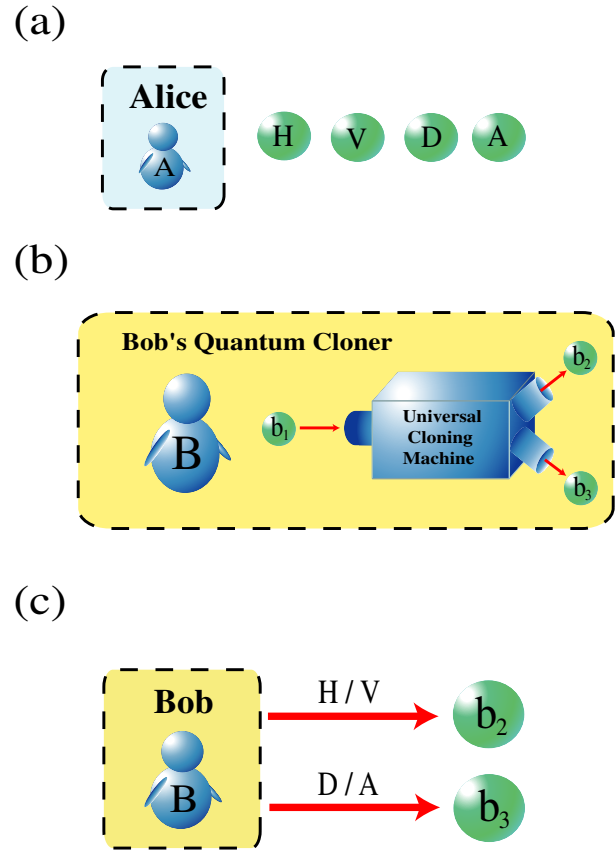}
\caption{The steps of the BB84 scheme using the no-switching
protocol. (a) Alice sends to Bob a random ensemble of photons
chosen from 4 possible polarizations. (b) Bob clones the incoming
photons from Alice using a universal quantum cloning machine. This
produces the two clones $b_{2}$ and $b_{3}$. (c) Bob then measures
one clone in the horizontal/vertical basis (H/V) and the other
clone in the diagonal/antidiagonal (D/A) basis. After
communicating classically with Alice, Bob can then discard those
times where he used the wrong basis.}\label{Figure_BB84_Cloning}
\end{center}
\end{figure}

Alice sends a random ensemble of linearly polarized photons to
Bob. These photons are from a set of 4 possible polarizations:
horizontal (H) $\equiv |0\rangle$, vertical (V) $\equiv
|1\rangle$, diagonal (D) $\equiv (|0\rangle + |1\rangle)/\sqrt2$
and anti-diagonal (A) $\equiv (|0\rangle - |1\rangle)/\sqrt2$. Bob
then uses a quantum cloning machine \cite{BH96} that takes one
input state and outputs two identical clones. Bob can therefore
measure the first set of clones with the rectilinear bases and the
other set with the diagonal bases. This guarantees that Bob always
measures in the correct basis. The classical steps of
reconciliation and privacy amplification follows. Figure
\ref{Figure_BB84_Cloning} gives an illustration of the protocol.
\subsection{Bob's Quantum Cloning Machine}

We assume that Bob has a universal quantum cloning machine at his
station. This enables him to clone with a maximum fidelity of
$5/6$ or $\approx83\%$ of the original input state. The universal
quantum cloning machine is state independent so will copy any
input state with the same fidelity. Therefore it is suited to QKD
as there is no prior knowledge of which of the four polarization
states Alice has sent.

To illustrate how the universal quantum cloner works we suppose
that Bob wants to copy an arbitrary pure state $|\psi_{1} \rangle
= \alpha |0\rangle + \beta |1\rangle$ (this pure state corresponds
to any one of the four photon polarizations). In our protocol
Bob's inputs a given qubit $b_{1}$ into his cloning machine which
outputs the identical clones $b_{2}$ and $b_{3}$. Here the state
of the cloning machine is given by $|C\rangle_{b}$. Bob wants to
copy the bases $|0\rangle$ and $|1\rangle$ of the arbitrary pure
state. This initial preparation by Bob of his cloning machine is
given by
\begin{eqnarray}\label{cloner_preparation1}
|0\rangle |0 \rangle |C\rangle_{b}\\\label{cloner_preparation2}
|1\rangle |0 \rangle |C\rangle_{b}
\end{eqnarray}
where $|0 \rangle$ is the blank state (e.g. the blank paper in
copying machines) that is used to make the clones with. The
resulting output of the cloning machine in
Eqs.~(\ref{cloner_preparation1},\ref{cloner_preparation2}) is
given by \cite{BH96}
\begin{eqnarray}
&\longrightarrow&  \sqrt{\frac{2}{3}}|00\rangle |\uparrow\rangle +
\sqrt{\frac{1}{6}} |+\rangle |\downarrow\rangle\\
&\longrightarrow&  \sqrt{\frac{2}{3}}|11\rangle |\downarrow\rangle
+ \sqrt{\frac{1}{6}} |+\rangle |\uparrow\rangle
\end{eqnarray}
where $|+\rangle = (|10\rangle) + |01\rangle)/\sqrt{2}$ and
$|\uparrow\rangle$ and $|\downarrow \rangle$ represent the output
states of the quantum cloning machine. The resulting clones are
identical, with each being of worse quality than the original
qubit that Bob used as the input state. We are then able to
perform a partial trace over the cloning machine states which
allows us to look at any subsystem of the original density
operator. After performing a partial trace over the cloning
machines we are left with a density operator describing the two
output clones $\hat{\rho}_{b_{2}b_{3}}^{out}$. To find the density
operator of either of the two clones individually, we can perform
a partial trace over either of the two clones. The resulting
reduced density operator can then be written as
\begin{eqnarray}\label{Cloner output}
\hat{\rho}_{b_{2}}^{out} &=& \frac{5}{6} |\psi_{1}\rangle \langle
\psi_{1}| + \frac{1}{6} |\psi_{2}\rangle \langle \psi_{2}|
\end{eqnarray}
where $|\psi_{1}\rangle = \alpha|0\rangle + \beta |1\rangle$ and
$|\psi_{2}\rangle = \beta|0\rangle - \alpha |1\rangle$.
Eq.(\ref{Cloner output}) is equal to the other density operator
$\hat{\rho}_{b_{3}}^{out}$ due to the symmetry of the system.
Eq.~(\ref{Cloner output}) tells us that the clones comprise of
$5/6$ of the original input state and $1/6$ of the errors of the
universal quantum cloning machine. Therefore by having these
clones of the original states sent by Alice, Bob can abandon
switching and measure both bases at the same time.

\subsection{Information Rate}

Finally to determine whether the no-switching protocol works in
the BB84 case we again need to consider information rates. Unlike
the continuous variable information rates we previously used, we
now need to use a (discrete) binary symmetric channel given by
\cite{Shannon}
\begin{eqnarray}\label{inforate_discrete}
I_{AB} = 1 + p_{e} {\rm log_{2}}p_{e} + (1 - p_{e}) {\rm
log_{2}}(1 - p_{e})
\end{eqnarray} \\\nonumber
where $p_{e}$ is the error probability between Alice and Bob's
mutual information. In the normal switching BB84 protocol (where
we neglect Eve for the moment), Alice and Bob have an error
probability of $25\%$. However after key sifting,where they
discard incorrect basis measurements, they have zero error
probability leading to $I_{AB} = 1$ bits/signal. But in actual
fact $I_{AB} = 0.5$ bits/signal as Alice and Bob have, on average,
thrown away half of their original data in the classical
communication step.

\subsection{To Switch Or Not To Switch}

In our no-switching protocol for the BB84 protocol we have an
error probability of $1/6$ or $\approx17\%$ (from Bob's cloning
machine), compared with an error rate of zero for the BB84
protocol, after Alice and Bob discard the results for the
incorrect measurement basis. Substituting this $17\%$ error into
Eq.(\ref{inforate_discrete}) gives an information rate of $I_{AB}
= 0.35$ bits/signal. This information rate is slower than the
$I_{AB} = 0.5$ bits/signal for the BB84 protocol and thus it is
more beneficial to use the original BB84 scheme. It is important
to note that the $17\%$ error is from Bob's station and not the
quantum channel (i.e. not from Eve's tampering). Normally an error
rate of $17\%$ from the quantum channel is too high and would mean
that Bob and Alice would not be able to distill a secure key. In
our case, before Bob starts cloning, he and Alice can test the
quantum channel for eavesdropping and provided it is below a safe
threshold, Bob can continue cloning. In any event randomly
switching the measurement bases when using discrete variable QKD
gives a higher information rate than using the no-switching
protocol.


\section{Conclusion}

Quantum key distribution has long made use of the random switching
of measurement bases to ensure the security of the protocol. We
have shown that switching is not a necessary requirement for
coherent state continuous variable quantum key distribution. This
was demonstrated via the no-switching protocol. The no-switching
protocol gives higher information rates than protocols that use
switching and also offers a simpler experimental setup.

We investigated a physical implementation of the eavesdropping
scheme in the form of  an entangling feed-forward attack. We
showed that this attack was no more effective than a simpler
coherent feed-forward attack that did not use entanglement. We
then conjectured that the coherent feed-forward attack was the
optimal attack assuming individual Gaussian attacks.

Finally we have shown that there is no advantage by applying the
no-switching protocol to the original BB84 protocol which employs
single photon states. This is an interesting result as it
highlights the differences between QKD with single photon states
compared to continuous variables.

We would like to thank T. Williams and D. Pulford for useful
discussions, and the financial support of the Australian Research
Council and the Australian Department of Defence.


\begin{thebibliography}{99}


\bibitem{Wiesner} S.Wiesner, SIGACT News {\bf 15}, 78 (1983).

\bibitem{BB84} C.H. Bennett and G.Brassard, in {\it Proceedings IEEE International Conference on Computers, Systems and Signal Proceedings
(Bangalore)} (IEEE, New York, 1984), pp. 175-179.

\bibitem{QKD} N. Gisin, G. Ribordy, W.Tittel, and H.Zbinden, Rev.
Mod. Phys. {\bf 74}, 145 (2002).

\bibitem{WZ82} W.K. Wooters and W.H. Zurek, {\it Nature} {\bf
299}, 802 (1982);

\bibitem{Mau93} U.M Maurer, IEEE Trans. Inf. Theory 39, 733 (1993); G. Brassard
and L. Salvail, Advances in Cryptology- EUROCRYPT93, Lecture Notes
Computer Science Vol. 765 (Springer-Verlag, Berlin, 1994), pp.
411– 423.

\bibitem{Ben95} C. H. Bennett et al., IEEE Trans. Inf. Theory 41, 1915
(1995).

\bibitem{Ver26} G.S. Vernam, J.Am.Inst. Electr. Eng. {\bf 45}, 109
(1926).

\bibitem{May98} D. Mayers, quant-ph/9802025, (1998).

\bibitem{Lo99} H.K. Lo and H.F. Chau, Science {\bf 283},
2050-2056, (1999).

\bibitem{Sho00} P.W. Shor and J.Preskill, {\it \prl} {\bf 85},
441-444, (2000).

\bibitem{Eke91} A.K. Ekert {\it \prl} {\bf 67}, 661
(1991).

\bibitem{Ben92} C.H. Bennett, {\it \prl}, {\bf 68} 3121 (1992).

\bibitem{ral00} T.C Ralph, Phys. Rev. A {\bf 61}, 010303(R) (1999).

\bibitem{Bra04} S.L. Braunstein and P.V. Loock, Rev. Mod. Phys. {\bf{77}},
513-577 (2005); quant-ph/0410100 (2004).

\bibitem{hil00} M.Hillery, Phys. Rev. A {\bf 61}, 022309 (2000).

\bibitem{Rei00} M.D. Reid, Phys. Rev. A {\bf 62}, 062308 (2000).

\bibitem{Cer01} N.J. Cerf, M. Levy, and G. Van Assche,
Phys. Rev. A {\bf 63}, 052311 (2001).

\bibitem{G&G3dB} F. Grosshans and Ph. Grangier, {\it \prl} {\bf 88}, 057902 (2002).

\bibitem{G&GReverse}F. Grosshans, and Ph. Grangier, quant-ph/0204127 (2002).

\bibitem{Sil02} C.Silberhorn, T.C. Ralph, N.Lutkenhaus, G.Leuchs,
{\it \prl} {\bf 89}, 167901 (2002).

\bibitem{Got01} D. Gottesman and J. Preskill, Phys. Rev. A {\bf 63}, 022309 (2001).

\bibitem{Ibl04} S. Iblisdir, G. Van Assche, and N. J. Cerf, {\it
\prl}. {\bf 93}, 170502 (2004).

\bibitem{Gro05} F. Grosshans , {\it \prl} {\bf
94}, 020504 (2005).

\bibitem{Gro03} F. Grosshans {\it et al}, Nature {\bf 421}, 238 (2003).

\bibitem{Lor04} S. Lorenz, N. Korolkova, G. Leuchs, {\it Appl. Phys.
B}, {\bf 79}, 273 (2004).

\bibitem{Lan05} A. Lance {\it et al}, quant-ph/0504004, (2005).

\bibitem{WLBS+04} C. Weedbrook {\it et al}, {\it \prl} {\bf 93}, 170504
(2004); quant-ph/0405105, (2004).

\bibitem{Fil05} R. Filip, L. Mista, and P. Marek, {\it Phys. Rev. A} {\bf 71}, 012323 (2005).

\bibitem{CRD04} P.T. Cochrane, T.C Ralph and A. Dolinska, Phys. Rev. A
{\bf 69}, 042313 (2004).

\bibitem{Shannon} C.E. Shannon, Bell Syst. Tech. J. {\bf 27}, 623
(1948).

\bibitem{Nie00} M.A. Nielsen and I.L. Chuang, {\it Quantum Computation and Quantum Information}
(Cambridge University Press, 2000).

\bibitem{GLP98} Grangier, Ph., Levenson, J.A. \& Poizat, J.-Ph. Quantum
non-demolition measurements in optics. {\it Nature} {\bf 396},
537-542 (1998).

\bibitem{Sak94} J.J. Sakurai, {\it Modern Quantum Mechanics}
(Addison-Wesley New York, 1994).

\bibitem{Ein35} A. Einstein, B. Podolsky, and N.Rosen, Phys. Rev.
{\bf 47}, 777 (1935).

\bibitem{BH96} V. Buzek and M.Hillery, Phys. Rev. A
{\bf 54}, 1844 (1996).

\bibitem{Bos02} K. Bostroem and T. Felbinger, {\it \prl} {\bf
89}, 187902 (2002).

\end{thebibliography}
\end{document}